\documentclass[aps,prl,twocolumn,superscriptaddress]{revtex4-1}
\usepackage{amsmath,amssymb,graphicx}

\begin{document}

\title{Optical Trapping and Control of a Dielectric Nanowire by a Nanoaperture}

\author{Mehdi Shafiei Aporvari}
\email{Corresponding author:mshphy@gmail.com}
\affiliation{Department of Physics, Faculty of Science, University of Isfahan, Hezar Jerib, 81746-73441, Isfahan, Iran}
\affiliation{Soft Matter Lab, Department of Physics, Bilkent University, Ankara 06800, Turkey}

\author{Fardin Kheirandish}
\affiliation{Department of Physics, Faculty of Science, University of Isfahan, Hezar Jerib, 81746-73441, Isfahan, Iran}

\author{Giovanni Volpe}
\affiliation{Soft Matter Lab, Department of Physics, Bilkent University, Ankara 06800, Turkey}
\affiliation{UNAM -- National Nanotechnology Research Center, Bilkent University, Ankara 06800, Turkey}
\date{\today}
\begin{abstract}
We demonstrate that a single sub-wavelength nanoaperture in a metallic thin film can be used to achieve dynamic optical trapping and control of a single dielectric nanowire. A nanoaperture can trap a nanowire, control its orientation when illuminated by a linearly-polarized incident field, and also rotate the nanowire when illuminated by a circularly-polarized incident field. Compared to other designs, this approach has the advantages of a low-power driving field entailing low heating and photodamage.
\end{abstract}

\maketitle

Optical manipulation and control of nanoparticles is potentially important in many areas of physical and life sciences \cite{marago2013optical}. In particular, controlling the position and orientation of elongated objects leads the way towards exciting applications: in nanotechnology, optically trapped semiconducting nanowires have been translated, rotated, cut and fused in order to realize complex nanostructures \cite{agarwal2005manipulation,pauzauskie2006optical,castelino2005manufacturing}; in spectroscopy, the composition and morphology of a sample have been probed by scanning optically trapped nanowires \cite{nakayama2007tunable} and polymer nanofibres \cite{neves2010rotational} over the sample's surface; in biophysics, many bacteria, viruses and macromolecules with rod-like shapes have been optically manipulated and studied \cite{capitanio2013interrogating}. However, optical manipulation of nanoobjects is particularly challenging; in fact, the techniques developed for optical manipulation of microparticles and, in particular, standard three-dimensional optical tweezers ---~i.e., tightly focused laser beams capable of confining microparticles \cite{ashkin1986observation,jones2015book}~--- cannot be straightforwardly scaled down to the nanoscale. This is mainly due to the fact that optical forces scale down with particle volume and are ultimately overwhelmed by the presence of thermal fluctuations \cite{marago2013optical}. 
Recently, several novel approaches have been put forward to overcome such limitations \cite{righini2008surface,juan2011plasmon,wang2011trapping,marago2013optical}. Thanks to the strong field enhancement associated with plasmonic resonances, plasmonic optical traps have been particularly successful in trapping ever smaller particles down to single molecules \cite{juan2011plasmon}. Furthermore, by altering the illumination conditions it has been possible to control the properties of the trap, e.g., tuning the strength of a plasmonic trap \cite{righini2008surface} and rotating an optically trapped particle around a plasmonic nanopillar \cite{wang2011trapping}. Also, hybrid plasmonic fields arising from the interaction between the surface plasmon polaritons (SPPs) along a metal surface and the localized surface plasmons (LSP) on a metallic nanowire lead to strong field enhancement and have been employed to trap and controllably orient a single gold nanowire \cite{zhang2014plasmonic}.  Finally, elongated metallic nanoparticles have also been aligned and rotated using conventional optical tweezers \cite{tong2009alignment}.
\begin{figure}[h!]
\centerline{\includegraphics[width=\columnwidth]{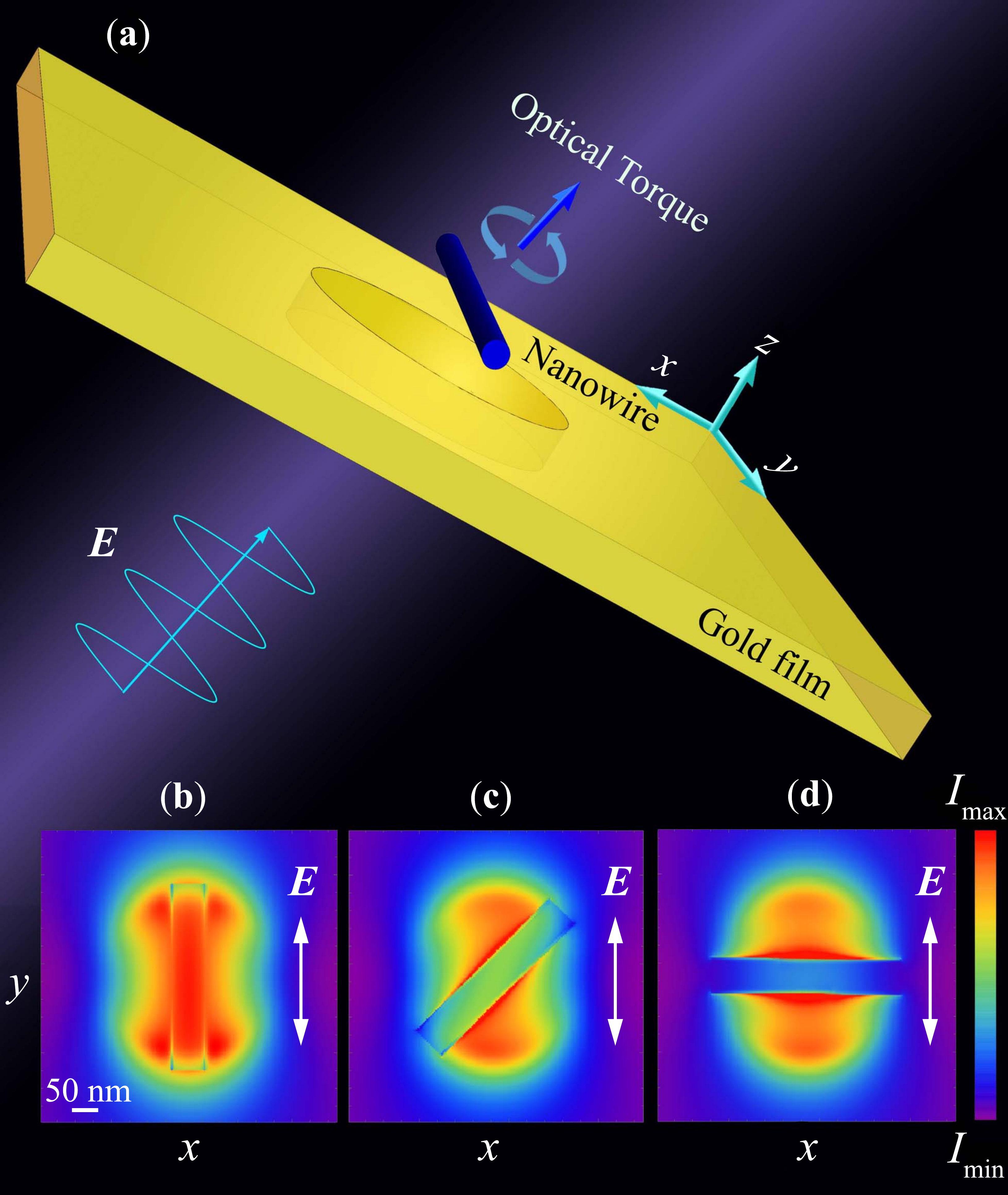}}
\caption{(color online) (\textbf{a}) Schematic of the configuration for the optical trapping and control of a nanowire by a nanoaperture.
(\textbf{b})-(\textbf{d}) Electric field intensity distribution in the $xy$-plane passing though the center of the nanowire (length $L = 350\,{\rm nm}$, radius $r = 30\,{\rm nm}$, medium refractive index $n_{\rm m} = 1.33$) for various orientations of the nanowire, i.e., (\textbf{b}) $\theta = 0$, (\textbf{c}) $\theta = \pi/4$ and (\textbf{d}) $\theta = \pi/2$. The nanoaperture radius is $155\,{\rm nm}$ in a 100-nm-thick gold film. The nanowire is placed $50\,{\rm nm}$ above the nanoaperture, so that the gap between the nanowire and the aperture is $20\,{\rm nm}$. The incident electric field is a plane wave linearly-polarized along the direction indicated by the arrows in (\textbf{b})-(\textbf{d}) with input intensity $I_0 = 10\,{\rm mW }\mu{\rm  m^{-2}}$.
}
\label{fig1}
\end{figure}

\begin{figure}[t!]
\centerline{\includegraphics[width=\columnwidth]{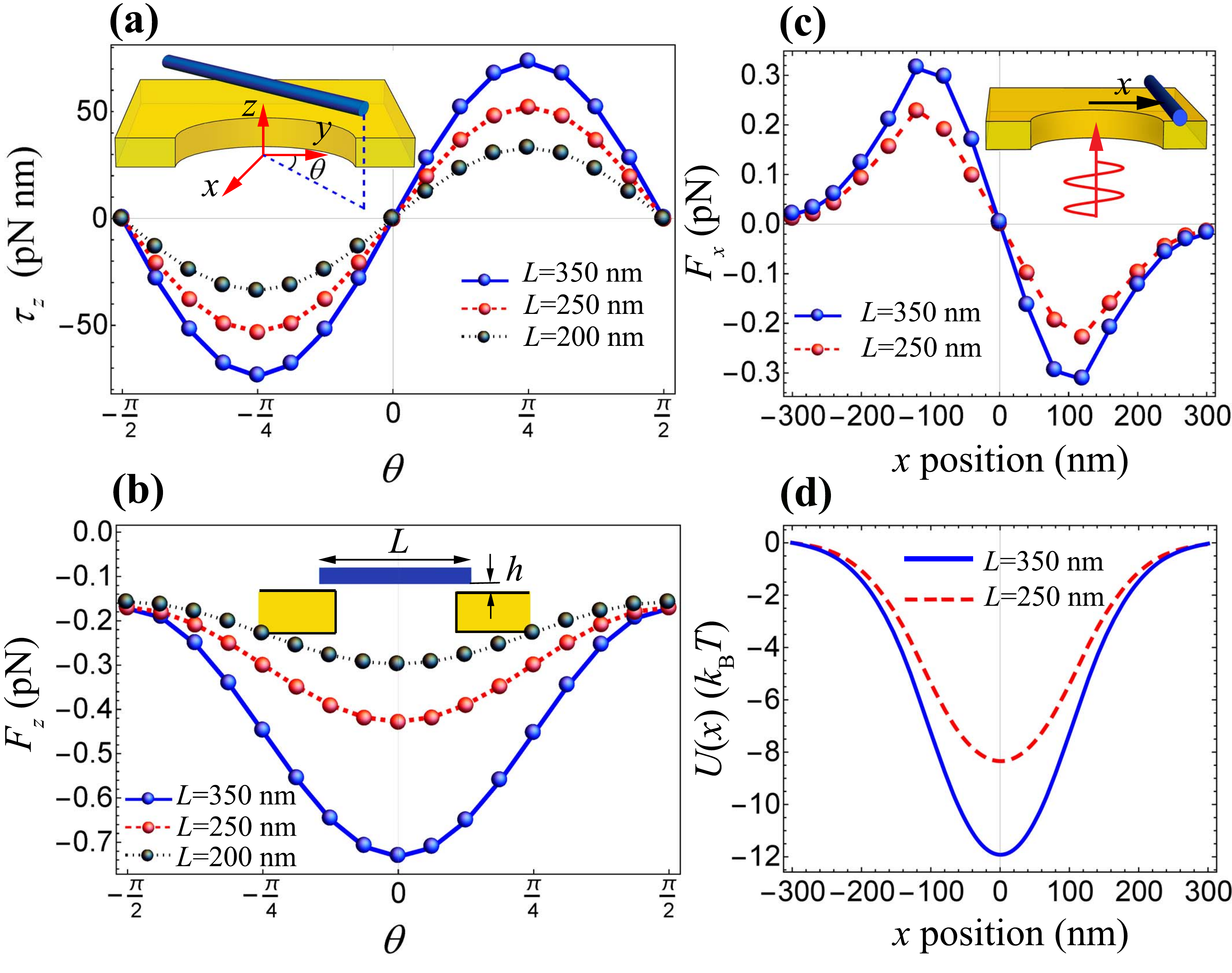}}
\caption{(color online) (\textbf{a}) Optical torque, $\tau_z$, and (\textbf{b}) normal optical force, $F_{z}$, on a nanowire of length $L=200$, $250$ and $300\,{\rm nm}$ trapped above a nanohole as a function of the angle $\theta$ between the nanowire and the incident field polarization (along the $y$-direction).
(\textbf{c}) Lateral optical restoring force, $F_x$, acting on nanowires of length $L=250$ and $350\,{\rm nm}$ and
(\textbf{d}) relative trapping potentials, $U(x)$.
In all cases, $r=30\,{\rm nm}$, $h=20\,{\rm nm}$ and $I_0=10\,{\rm mW}\mu{\rm m^{-2}}$. }
\label{fig2}
\end{figure} 

Amongst the various metallic nanostructures that have been employed for plasmonic optical manipulation, subwavelength apertures in thin metal films have generated a considerable amount of interest \cite{garcia2010light,genet2007light,lezec2002beaming,juan2009self}.
Extraordinary optical transmission was first observed on arrays of nanoholes in metallic films, where it was due to the coupling of light to SSPs excited in the periodically patterned metal film \cite{ebbesen1998extraordinary}.
Later, it was demonstrated that, even for a single nanohole in a flat metal surface, the excitation of LSPs on the aperture ridge can alter its transmission properties \cite{degiron2004optical}; in particular, the electric field intensity pattern was shown to display two high intensity spots parallel to the polarization direction of the incident light and attributed to a dipole moment oriented in the plane of the metal film and parallel to the axis joining the spots. 
One of the main advantages of using sub-wavelength nanoholes for optical trapping is that they permit one to reduce the required local intensity compared with conventional trapping, therefore significantly reducing the likelihood of optical damage \cite{juan2011plasmon}.

The main idea of the work presented in this Letter is to use the intensity distribution with two hotspots generated at a nanoaperture as a handle to trap and control elongated objects such as nanowires. We show that nanoapertures can be efficiently used to trap dielectric nanowires with low power and to control their orientation and movement through the polarization of the input beam. Compared to previous designs, the approach we propose has the advantage of a low-power driving field and an extremely simple design. 

As shown in Fig.~\ref{fig1}a, we consider a dielectric nanowire with radius $r$ and length $L$ placed at a distance $h$ above a cylindrical nanoaperture of radius $155\,{\rm nm}$ in a 100-nm-thick gold film. The refractive index of the nanowire is chosen as $2.0$, which is near to the index of ZnO nanowires \citep{pauzauskie2006optical}.  A linearly-polarized plane wave with wavelength $1064\,{\rm nm}$ illuminates the sample from the side opposite to the one where the nanowire lies. The long axis of the nanowire lies on the $xy$-plane at an angle $\theta$ with respect to the polarization direction. The medium is water (refractive index $n_{\rm m} = 1.33$). We compute the electromagnetic field using a home-made software based on a three-dimensional finite-difference time-domain (3D-FDTD) algorithm together with a total-field/scatter-field technique and employing a small grid size (which depends on the specific configuration, but is always less than $3\,{\rm nm}$) in order to account for all plasmonic near-field behaviors \cite{taflove2005computational}. We bound the simulation domain with a convolutional perfectly matched layer (CPML) \cite{taflove2005computational}. For the dielectric constant of the film, we use the Drude model with parameters that fit the experimental values of the dielectric data for gold \cite{johnson1972optical}. Figs.~\ref{fig1}b-d show the intensity distributions in the $xy$-plane passing through the center of a nanowire placed at a distance of $50\,{\rm nm}$ above the metal film (so that the gap between the nanowire and the metallic surface is $h=20\,{\rm nm}$) with various orientations. The resulting trapping is mainly due to the creation of light gradients around the aperture and the self-induced back-action (SIBA) mechanism \cite{juan2009self}. In fact, the nanoaperture (radius $155\,{\rm nm}$) is tuned near the cutoff of the resonance (at $165\,{\rm nm}$ when illuminated at $1064\,{\rm nm}$) so that the presence of a particle with high refractive index increases the effective aperture size and, therefore, its transmission and associated optical forces \cite{juan2009self}. There are two hotspots on the rim of the nanohole along the incident field polarization direction. As we will see in the following, the nanowire tends to align itself along the direction of these hotspots and this effect can be exploited to control the nanowire's orientation.

Once the electromagnetic fields have been calculated using 3D-FDTD, we proceed to calculate the optical forces and torques acting on the nanowire using the Maxwell stress tensor (MST) method \cite{borghese2007scattering}. The time-averaged force and torque acting on the center of mass of the nanowire are
\begin{equation}
\left<\textbf{F}\right>=\int_{S} \left<\mathbb{T}\right> \cdot \hat{\textbf{n}} \; dS
\end{equation}
and
\begin{equation}
\left<\boldsymbol{\tau}\right>=-\int_{S} \left<\mathbb{T}\right> \times \textbf{r} \cdot \hat{\textbf{n}} \; dS \; ,
\end{equation}
where $S$ is a surface enclosing the nanowire, $\hat{\textbf{n}}$ is the unit vector perpendicular to the surface, $\textbf{r}$ is the position of the surface element, and $\left<\mathbb{T}\right>$ is the time-averaged Maxwell stress tensor for harmonic fields, i.e.,
\begin{equation}
\left<\mathbb{T}\right>
=
\frac{1}{2} \Re 
\left\{
 \epsilon \textbf{E}\textbf{ E}^{*} +
 \mu \textbf{H}\textbf{ H}^{*} -
 \frac{\textbf{I}}{2} \left( \epsilon |\textbf{E}|^2 + \mu |\textbf{H}|^2 \right)
\right\} \; ,
\end{equation}
where $\textbf{E}$ is the electric field, $\textbf{H}$ is the magnetic field, and $\epsilon$ and $\mu$ are the permittivity and permeability of the surrounding medium. A convergence analysis is performed for each configuration to ensure a correct force calculation.

\begin{figure}[t]
\centerline{\includegraphics[width=1\columnwidth]{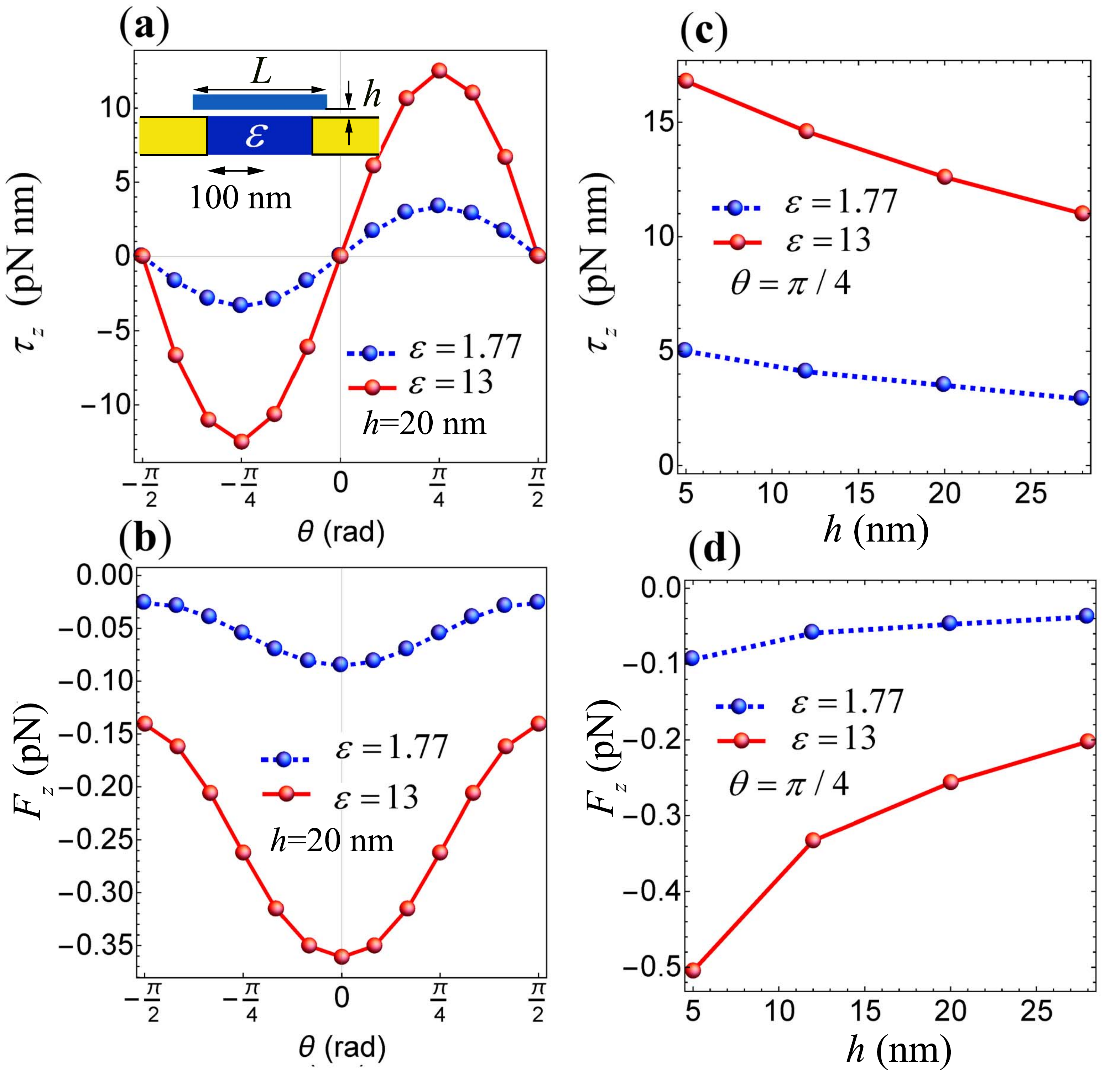}}
\caption{(color online) (\textbf{a},\textbf{c}) Optical torque and (\textbf{b},\textbf{d}) normal optical force on a nanowire ($L = 250\,{\rm nm}$, $r=20\,{\rm nm}$) placed on the exit side of a nanohole of radius $100\,{\rm nm}$ filled with water (dashed lines, $\epsilon=1.77 $) and silicon (solid lines, $\epsilon=13$) as a function (\textbf{a},\textbf{b}) of the angle $\theta$ between the nanowire and the incident field polarization direction and (\textbf{c},\textbf{d}) of the gap $h$ between the nanowire and the metallic layer. The incident intensity is $I_0 = 10\,{\rm mW}\mu{\rm  m^{-2}}$.}
\label{fig3}
\end{figure}

Fig.~\ref{fig2}a shows the torque acting on the nanowire along the $z$-axis as a function of its orientation $\theta$ with respect to the incident field polarization direction. The torque magnitude is maximum at $\theta = \pm \pi/4$ and zero at $\theta = 0$ (stable equilibrium) and $\pm\pi/2$ (unstable equilibrium). Therefore, the nanowire tends to align itself with the polarization of the incident electromagnetic field.
In a simplified picture, which nevertheless permits one to gain some qualitative insight into the nanowire orientation process, this behavior is similar to the alignment of an induced dipole with the polarization direction of a monochromatic electromagnetic field.
In fact, since the external electric field induces a dipole moment ${\bf p}=\pmb{\alpha}\cdot{\bf E}$, where $\pmb{\alpha}$ is the polarizability tensor, the electrostatic torque exerted on the particle is $\pmb{\tau}={1\over2}\Re\{\pmb{p}\times {\bf E}^* \} $. Considering a symmetric ellipsoid, this torque is
$\tau_z=\alpha_{\rm eff}{E}^2\sin(2\theta)$,
where $E=|\pmb{E}|$, $\alpha_{\rm eff}=\Re\{ \alpha_{\rm long}-\alpha_{\rm trans}\}$, $\alpha_{\rm long}$ and $\alpha_{\rm trans}$ are the polarizabilities along the longitudinal and transverse axes of the ellipsoid \cite{washizu1992orientation}. This equation shows that the torque has a sinusoidal behavior with angular period $\pi$, which is consistent with the results shown in Fig.~\ref{fig2}a. The torque for $\theta=\pi/2$ increases as the length of the nanowire increases up to a maximum for $L=350\,{\rm nm}$, i.e., when the length of the nanowire is slightly larger than the diameter of the nanohole; a further increase of the nanowire length makes its sides go outside the LSP associated to the nanohole and, thus, does not contribute to the overall optical torque. The torque for $\theta=0$ is negligible for nanowires with lengths between $L=80$ and $600\,{\rm nm}$. 

Fig.~\ref{fig2}b shows the normal force $(F_z)$, whose negative sign indicates that this is a restoring force pulling the nanowire toward the nanoaperture, where the fields are more enhanced resulting in even stronger transverse and normal trapping forces. We remark that the nanowire also undergoes Brownian motion so that it does not stick to the nanoaperture, but keeps on jiggling above it; in fact, Brownian motion prevents sticking the nanowire to the surface making  plasmonic trapping and control practically possible \citep{zhang2014plasmonic,wang2009propulsion,min2013focused} and, in fact, determines the characteristic trapping time \citep{juan2009self}. The nanohole confines the nanowire also along the transverse $x$-direction, as can be seen from the forces and associated potentials (Figs.~\ref{fig2}c and \ref{fig2}d). Importantly, the depth of the potential is sufficient to ensure stable trapping, as it is $12\,k_{\rm B}T$ ($8\,k_{\rm B}T$) at $I_0 = 10\,{\rm mW}\mu{\rm  m^{-2}}$ for $L=350\,{\rm nm}$ ($250\,{\rm nm}$). Furthermore, the transmission properties of the nanoaperture are very sensitive to its local environment and, thus, to the presence of the nanowire. For example, for the configuration presented in Fig.~\ref{fig1}, the presence of the nanowire increases the transmitted intensity by $8\%$ ($6\%$) for a nanowire of length $L=350\,{\rm nm}$ ($250\,{\rm nm}$); this change is of the same order of magnitude of what is observed and calculated in Ref.~\citenum{juan2009self}. Importantly, this intensity change can be experimentally monitored and can be used to detect the presence of an optically trapped nanowire.

It is also interesting to consider the case when the nanoaperture is filled with a high-index dielectric material. The use of such material, instead of water, in addition to confining the nanowire on a flat surface, gives the additional advantage of improving the nanoaperture transmission and, thus, of narrowing the transmission resonance \cite{garcia2002light}. Noting that filling the nanohole shifts the transmission resonance towards a smaller radius/wavelength ratio, we consider a nanoaperture with radius $100\,{\rm nm}$ in a gold film of thickness $100\,{\rm nm}$, i.e., an aperture similar to the one in Fig.~\ref{fig2} but with smaller radius. This structure is tuned close to the second peak in the transmission spectrum. As shown in Fig.~\ref{fig3}a, for such a nanoaperture filled with silicon ($\epsilon=13$), the optical torque on a nanowire ($L=250\,{\rm nm}$, $r=20\,{\rm nm}$) is enhanced nearly three times compared with the undressed hole. Also the normal optical force is considerably enhanced, as can be seen in Fig.~\ref{fig3}b. Both torque (Fig.~\ref{fig3}c) and force (Fig.~\ref{fig3}d) increase as the gap $h$ between the nanowire and the surface decreases. Therefore, filling nanoapertures with high-index dielectric material can improve the capabilities of the structure and make it more appropriate to efficiently trap and manipulate even smaller particles.

\begin{figure}[t]
\centerline{\includegraphics[width=1\columnwidth]{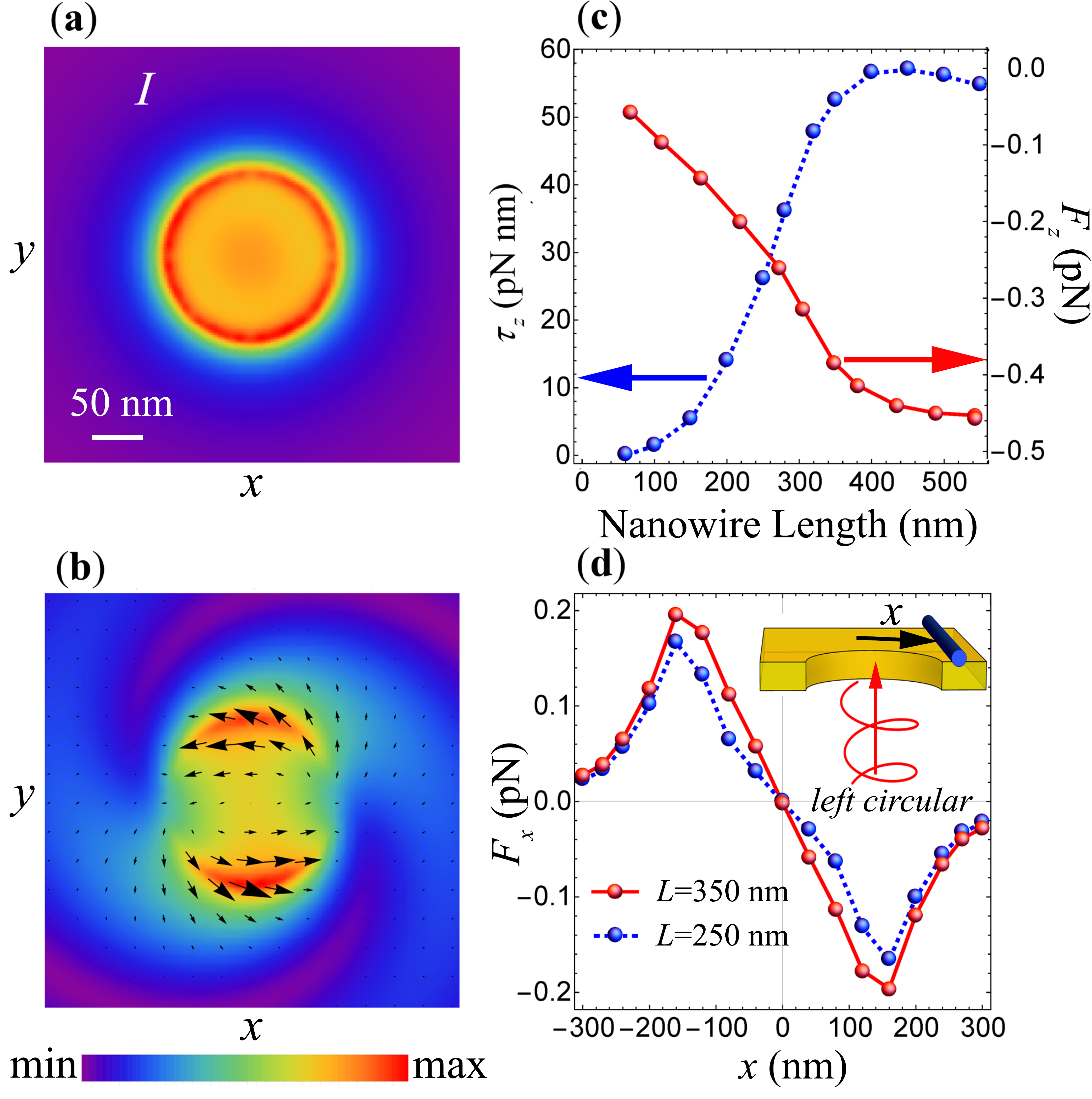}}
\caption{(color online) (\textbf{a}) Electric field intensity distribution and (\textbf{b}) instantaneous electric field magnitude distribution (shades) and Poynting vector (arrows) in the $xy$-plane $20\,{\rm nm}$ above the nanoaperture for left circularly-polarized incident light. 
(\textbf{c}) Optical torque, $\tau_z$, and normal optical force,$F_z$, for left circularly-polarized incident light as a function of nanowire length $L$. 
(\textbf{d}) Lateral optical force, $F_x$, acting on the nanowire for left circularly-polarized incident light for nanowires with lengths $L = 250$ and $350\,{\rm nm}$.
In (\textbf{c}) and (\textbf{d}), $h = 20\,{\rm nm}$ and $ r=30\,{\rm nm}$. The incident intensity is $I_0 = 10\,{\rm mW}\mu{\rm  m^{-2}}$.}
\label{fig4}
\end{figure}

Finally, we also consider the case of circularly-polarized illumination. In this case, differently from the case of a linearly-polarized incident wave, the average field intensity is ring-shaped near the edge of the nanoaperture, as shown in Fig.~\ref{fig4}a. In fact, the two hotspots generated by the instantaneous electric fields (Fig.~\ref{fig4}b) rotate around the edge of the nanoaperture at the frequency of the incident field, leading to a transverse clockwise or counterclockwise energy flow depending on whether the incident wave is left or right circularly-polarized. As can be seen in Figs.~\ref{fig4}a and \ref{fig4}b, a plasmonic optical vortex is generated and, thus, the nanoaperture can be thought as a device capable of converting the spin angular momentum (SAM) of the circularly-polarized incident beam to orbital angular momentum (OAM) of the evanescent field \cite{gorodetski2013generating}.
Fig.~\ref{fig4}c shows the optical torque (solid line) and normal force (dashed line) acting on a nanowire placed on the exit side of the aperture for left-circularly-polarized illumination. Importantly, although the optical torque and normal force magnitudes are a slightly smaller than the highest corresponding torque and force in the case of  linear polarization case (cfr. Fig.~\ref{fig2}a-b), they are independent of the nanowire orientation. 
Fig.~\ref{fig4}d shows the lateral optical force on the nanowire for $L=250$ and $350\,{\rm nm}$ as a function of their lateral position along the $x$-direction. The nanowires experience the largest restoring force when moving towards the edges of the nanoaperture, while the central position corresponds to a stable equilibrium.

In conclusion, we have demonstrated that it is possible to use a nanoaperture to trap a nanowire and to control its orientation with very low incident power. Furthermore, we have shown that employing circularly-polarized light it is possible to rotate the nanowire at a constant rotation rate. The simplicity of the nanoaperture geometry will permit easy fabrication of this nanodevice.

GV acknowledges funding from Marie Curie Career Integration Grant (MC-CIG) PCIG11GA-2012-321726 and
a Distinguished Young Scientist award of the Turkish Academy of Sciences (T\"UBA).
\bibliography{biblio}

\begin{thebibliography}{28}%
\makeatletter
\providecommand \@ifxundefined [1]{%
 \@ifx{#1\undefined}
}%
\providecommand \@ifnum [1]{%
 \ifnum #1\expandafter \@firstoftwo
 \else \expandafter \@secondoftwo
 \fi
}%
\providecommand \@ifx [1]{%
 \ifx #1\expandafter \@firstoftwo
 \else \expandafter \@secondoftwo
 \fi
}%
\providecommand \natexlab [1]{#1}%
\providecommand \enquote  [1]{``#1''}%
\providecommand \bibnamefont  [1]{#1}%
\providecommand \bibfnamefont [1]{#1}%
\providecommand \citenamefont [1]{#1}%
\providecommand \href@noop [0]{\@secondoftwo}%
\providecommand \href [0]{\begingroup \@sanitize@url \@href}%
\providecommand \@href[1]{\@@startlink{#1}\@@href}%
\providecommand \@@href[1]{\endgroup#1\@@endlink}%
\providecommand \@sanitize@url [0]{\catcode `\\12\catcode `\$12\catcode
  `\&12\catcode `\#12\catcode `\^12\catcode `\_12\catcode `\%12\relax}%
\providecommand \@@startlink[1]{}%
\providecommand \@@endlink[0]{}%
\providecommand \url  [0]{\begingroup\@sanitize@url \@url }%
\providecommand \@url [1]{\endgroup\@href {#1}{\urlprefix }}%
\providecommand \urlprefix  [0]{URL }%
\providecommand \Eprint [0]{\href }%
\providecommand \doibase [0]{http://dx.doi.org/}%
\providecommand \selectlanguage [0]{\@gobble}%
\providecommand \bibinfo  [0]{\@secondoftwo}%
\providecommand \bibfield  [0]{\@secondoftwo}%
\providecommand \translation [1]{[#1]}%
\providecommand \BibitemOpen [0]{}%
\providecommand \bibitemStop [0]{}%
\providecommand \bibitemNoStop [0]{.\EOS\space}%
\providecommand \EOS [0]{\spacefactor3000\relax}%
\providecommand \BibitemShut  [1]{\csname bibitem#1\endcsname}%
\let\auto@bib@innerbib\@empty
\bibitem [{\citenamefont {Marag{\`o}}\ \emph {et~al.}(2013)\citenamefont
  {Marag{\`o}}, \citenamefont {Jones}, \citenamefont {Gucciardi}, \citenamefont
  {Volpe},\ and\ \citenamefont {Ferrari}}]{marago2013optical}%
  \BibitemOpen
  \bibfield  {author} {\bibinfo {author} {\bibfnamefont {O.~M.}\ \bibnamefont
  {Marag{\`o}}}, \bibinfo {author} {\bibfnamefont {P.~H.}\ \bibnamefont
  {Jones}}, \bibinfo {author} {\bibfnamefont {P.~G.}\ \bibnamefont
  {Gucciardi}}, \bibinfo {author} {\bibfnamefont {G.}~\bibnamefont {Volpe}}, \
  and\ \bibinfo {author} {\bibfnamefont {A.~C.}\ \bibnamefont {Ferrari}},\
  }\href@noop {} {\bibfield  {journal} {\bibinfo  {journal} {Nature Nanotech.}\
  }\textbf {\bibinfo {volume} {8}},\ \bibinfo {pages} {807} (\bibinfo {year}
  {2013})}\BibitemShut {NoStop}%
\bibitem [{\citenamefont {Agarwal}\ \emph {et~al.}(2005)\citenamefont
  {Agarwal}, \citenamefont {Ladavac}, \citenamefont {Roichman}, \citenamefont
  {Yu}, \citenamefont {Lieber},\ and\ \citenamefont
  {Grier}}]{agarwal2005manipulation}%
  \BibitemOpen
  \bibfield  {author} {\bibinfo {author} {\bibfnamefont {R.}~\bibnamefont
  {Agarwal}}, \bibinfo {author} {\bibfnamefont {K.}~\bibnamefont {Ladavac}},
  \bibinfo {author} {\bibfnamefont {Y.}~\bibnamefont {Roichman}}, \bibinfo
  {author} {\bibfnamefont {G.}~\bibnamefont {Yu}}, \bibinfo {author}
  {\bibfnamefont {C.}~\bibnamefont {Lieber}}, \ and\ \bibinfo {author}
  {\bibfnamefont {D.}~\bibnamefont {Grier}},\ }\href@noop {} {\bibfield
  {journal} {\bibinfo  {journal} {Opt. Express}\ }\textbf {\bibinfo {volume}
  {13}},\ \bibinfo {pages} {8906} (\bibinfo {year} {2005})}\BibitemShut
  {NoStop}%
\bibitem [{\citenamefont {Pauzauskie}\ \emph {et~al.}(2006)\citenamefont
  {Pauzauskie}, \citenamefont {Radenovic}, \citenamefont {Trepagnier},
  \citenamefont {Shroff}, \citenamefont {Yang},\ and\ \citenamefont
  {Liphardt}}]{pauzauskie2006optical}%
  \BibitemOpen
  \bibfield  {author} {\bibinfo {author} {\bibfnamefont {P.~J.}\ \bibnamefont
  {Pauzauskie}}, \bibinfo {author} {\bibfnamefont {A.}~\bibnamefont
  {Radenovic}}, \bibinfo {author} {\bibfnamefont {E.}~\bibnamefont
  {Trepagnier}}, \bibinfo {author} {\bibfnamefont {H.}~\bibnamefont {Shroff}},
  \bibinfo {author} {\bibfnamefont {P.}~\bibnamefont {Yang}}, \ and\ \bibinfo
  {author} {\bibfnamefont {J.}~\bibnamefont {Liphardt}},\ }\href@noop {}
  {\bibfield  {journal} {\bibinfo  {journal} {Nature Materials}\ }\textbf
  {\bibinfo {volume} {5}},\ \bibinfo {pages} {97} (\bibinfo {year}
  {2006})}\BibitemShut {NoStop}%
\bibitem [{\citenamefont {Castelino}\ \emph {et~al.}(2005)\citenamefont
  {Castelino}, \citenamefont {Satyanarayana},\ and\ \citenamefont
  {Sitti}}]{castelino2005manufacturing}%
  \BibitemOpen
  \bibfield  {author} {\bibinfo {author} {\bibfnamefont {K.}~\bibnamefont
  {Castelino}}, \bibinfo {author} {\bibfnamefont {S.}~\bibnamefont
  {Satyanarayana}}, \ and\ \bibinfo {author} {\bibfnamefont {M.}~\bibnamefont
  {Sitti}},\ }\href@noop {} {\bibfield  {journal} {\bibinfo  {journal}
  {Robotica}\ }\textbf {\bibinfo {volume} {23}},\ \bibinfo {pages} {435}
  (\bibinfo {year} {2005})}\BibitemShut {NoStop}%
\bibitem [{\citenamefont {Nakayama}\ \emph {et~al.}(2007)\citenamefont
  {Nakayama}, \citenamefont {Pauzauskie}, \citenamefont {Radenovic},
  \citenamefont {Onorato}, \citenamefont {Saykally}, \citenamefont {Liphardt},\
  and\ \citenamefont {Yang}}]{nakayama2007tunable}%
  \BibitemOpen
  \bibfield  {author} {\bibinfo {author} {\bibfnamefont {Y.}~\bibnamefont
  {Nakayama}}, \bibinfo {author} {\bibfnamefont {P.~J.}\ \bibnamefont
  {Pauzauskie}}, \bibinfo {author} {\bibfnamefont {A.}~\bibnamefont
  {Radenovic}}, \bibinfo {author} {\bibfnamefont {R.~M.}\ \bibnamefont
  {Onorato}}, \bibinfo {author} {\bibfnamefont {R.~J.}\ \bibnamefont
  {Saykally}}, \bibinfo {author} {\bibfnamefont {J.}~\bibnamefont {Liphardt}},
  \ and\ \bibinfo {author} {\bibfnamefont {P.}~\bibnamefont {Yang}},\
  }\href@noop {} {\bibfield  {journal} {\bibinfo  {journal} {Nature}\ }\textbf
  {\bibinfo {volume} {447}},\ \bibinfo {pages} {1098} (\bibinfo {year}
  {2007})}\BibitemShut {NoStop}%
\bibitem [{\citenamefont {Neves}\ \emph {et~al.}(2010)\citenamefont {Neves},
  \citenamefont {Camposeo}, \citenamefont {Pagliara}, \citenamefont {Saija},
  \citenamefont {Borghese}, \citenamefont {Denti}, \citenamefont {Iat{\`\i}},
  \citenamefont {Cingolani}, \citenamefont {Marag{\`o}},\ and\ \citenamefont
  {Pisignano}}]{neves2010rotational}%
  \BibitemOpen
  \bibfield  {author} {\bibinfo {author} {\bibfnamefont {A.~A.~R.}\
  \bibnamefont {Neves}}, \bibinfo {author} {\bibfnamefont {A.}~\bibnamefont
  {Camposeo}}, \bibinfo {author} {\bibfnamefont {S.}~\bibnamefont {Pagliara}},
  \bibinfo {author} {\bibfnamefont {R.}~\bibnamefont {Saija}}, \bibinfo
  {author} {\bibfnamefont {F.}~\bibnamefont {Borghese}}, \bibinfo {author}
  {\bibfnamefont {P.}~\bibnamefont {Denti}}, \bibinfo {author} {\bibfnamefont
  {M.~A.}\ \bibnamefont {Iat{\`\i}}}, \bibinfo {author} {\bibfnamefont
  {R.}~\bibnamefont {Cingolani}}, \bibinfo {author} {\bibfnamefont {O.~M.}\
  \bibnamefont {Marag{\`o}}}, \ and\ \bibinfo {author} {\bibfnamefont
  {D.}~\bibnamefont {Pisignano}},\ }\href@noop {} {\bibfield  {journal}
  {\bibinfo  {journal} {Opt. Express}\ }\textbf {\bibinfo {volume} {18}},\
  \bibinfo {pages} {822} (\bibinfo {year} {2010})}\BibitemShut {NoStop}%
\bibitem [{\citenamefont {Capitanio}\ and\ \citenamefont
  {Pavone}(2013)}]{capitanio2013interrogating}%
  \BibitemOpen
  \bibfield  {author} {\bibinfo {author} {\bibfnamefont {M.}~\bibnamefont
  {Capitanio}}\ and\ \bibinfo {author} {\bibfnamefont {F.~S.}\ \bibnamefont
  {Pavone}},\ }\href@noop {} {\bibfield  {journal} {\bibinfo  {journal}
  {Biophys. J.}\ }\textbf {\bibinfo {volume} {105}},\ \bibinfo {pages} {1293}
  (\bibinfo {year} {2013})}\BibitemShut {NoStop}%
\bibitem [{\citenamefont {Ashkin}\ \emph {et~al.}(1986)\citenamefont {Ashkin},
  \citenamefont {Dziedzic}, \citenamefont {Bjorkholm},\ and\ \citenamefont
  {Chu}}]{ashkin1986observation}%
  \BibitemOpen
  \bibfield  {author} {\bibinfo {author} {\bibfnamefont {A.}~\bibnamefont
  {Ashkin}}, \bibinfo {author} {\bibfnamefont {J.~M.}\ \bibnamefont
  {Dziedzic}}, \bibinfo {author} {\bibfnamefont {J.~E.}\ \bibnamefont
  {Bjorkholm}}, \ and\ \bibinfo {author} {\bibfnamefont {S.}~\bibnamefont
  {Chu}},\ }\href@noop {} {\bibfield  {journal} {\bibinfo  {journal} {Opt.
  Lett.}\ }\textbf {\bibinfo {volume} {11}},\ \bibinfo {pages} {288} (\bibinfo
  {year} {1986})}\BibitemShut {NoStop}%
\bibitem [{\citenamefont {Jones}\ \emph {et~al.}(2015)\citenamefont {Jones},
  \citenamefont {Marag\'{o}},\ and\ \citenamefont {Volpe}}]{jones2015book}%
  \BibitemOpen
  \bibfield  {author} {\bibinfo {author} {\bibfnamefont {P.~H.}\ \bibnamefont
  {Jones}}, \bibinfo {author} {\bibfnamefont {O.~M.}\ \bibnamefont
  {Marag\'{o}}}, \ and\ \bibinfo {author} {\bibfnamefont {G.}~\bibnamefont
  {Volpe}},\ }\href@noop {} {\emph {\bibinfo {title} {Optical tweezers:
  Principles and applications}}}\ (\bibinfo  {publisher} {Cambridge University
  Press},\ \bibinfo {year} {2015})\BibitemShut {NoStop}%
\bibitem [{\citenamefont {Righini}\ \emph {et~al.}(2008)\citenamefont
  {Righini}, \citenamefont {Volpe}, \citenamefont {Girard}, \citenamefont
  {Petrov},\ and\ \citenamefont {Quidant}}]{righini2008surface}%
  \BibitemOpen
  \bibfield  {author} {\bibinfo {author} {\bibfnamefont {M.}~\bibnamefont
  {Righini}}, \bibinfo {author} {\bibfnamefont {G.}~\bibnamefont {Volpe}},
  \bibinfo {author} {\bibfnamefont {C.}~\bibnamefont {Girard}}, \bibinfo
  {author} {\bibfnamefont {D.}~\bibnamefont {Petrov}}, \ and\ \bibinfo {author}
  {\bibfnamefont {R.}~\bibnamefont {Quidant}},\ }\href@noop {} {\bibfield
  {journal} {\bibinfo  {journal} {Phys. Rev. Lett.}\ }\textbf {\bibinfo
  {volume} {100}},\ \bibinfo {pages} {186804} (\bibinfo {year}
  {2008})}\BibitemShut {NoStop}%
\bibitem [{\citenamefont {Juan}\ \emph {et~al.}(2011)\citenamefont {Juan},
  \citenamefont {Righini},\ and\ \citenamefont {Quidant}}]{juan2011plasmon}%
  \BibitemOpen
  \bibfield  {author} {\bibinfo {author} {\bibfnamefont {M.~L.}\ \bibnamefont
  {Juan}}, \bibinfo {author} {\bibfnamefont {M.}~\bibnamefont {Righini}}, \
  and\ \bibinfo {author} {\bibfnamefont {R.}~\bibnamefont {Quidant}},\
  }\href@noop {} {\bibfield  {journal} {\bibinfo  {journal} {Nature Photon.}\
  }\textbf {\bibinfo {volume} {5}},\ \bibinfo {pages} {349} (\bibinfo {year}
  {2011})}\BibitemShut {NoStop}%
\bibitem [{\citenamefont {Wang}\ \emph {et~al.}(2011)\citenamefont {Wang},
  \citenamefont {Schonbrun}, \citenamefont {Steinvurzel},\ and\ \citenamefont
  {Crozier}}]{wang2011trapping}%
  \BibitemOpen
  \bibfield  {author} {\bibinfo {author} {\bibfnamefont {K.}~\bibnamefont
  {Wang}}, \bibinfo {author} {\bibfnamefont {E.}~\bibnamefont {Schonbrun}},
  \bibinfo {author} {\bibfnamefont {P.}~\bibnamefont {Steinvurzel}}, \ and\
  \bibinfo {author} {\bibfnamefont {K.~B.}\ \bibnamefont {Crozier}},\
  }\href@noop {} {\bibfield  {journal} {\bibinfo  {journal} {Nature Commun.}\
  }\textbf {\bibinfo {volume} {2}},\ \bibinfo {pages} {469} (\bibinfo {year}
  {2011})}\BibitemShut {NoStop}%
\bibitem [{\citenamefont {Zhang}\ \emph {et~al.}(2014)\citenamefont {Zhang},
  \citenamefont {Wang}, \citenamefont {Shen}, \citenamefont {Man},
  \citenamefont {Shi}, \citenamefont {Min}, \citenamefont {Yuan}, \citenamefont
  {Zhu}, \citenamefont {Urbach},\ and\ \citenamefont
  {Yuan}}]{zhang2014plasmonic}%
  \BibitemOpen
  \bibfield  {author} {\bibinfo {author} {\bibfnamefont {Y.}~\bibnamefont
  {Zhang}}, \bibinfo {author} {\bibfnamefont {J.}~\bibnamefont {Wang}},
  \bibinfo {author} {\bibfnamefont {J.}~\bibnamefont {Shen}}, \bibinfo {author}
  {\bibfnamefont {Z.}~\bibnamefont {Man}}, \bibinfo {author} {\bibfnamefont
  {W.}~\bibnamefont {Shi}}, \bibinfo {author} {\bibfnamefont {C.}~\bibnamefont
  {Min}}, \bibinfo {author} {\bibfnamefont {G.}~\bibnamefont {Yuan}}, \bibinfo
  {author} {\bibfnamefont {S.}~\bibnamefont {Zhu}}, \bibinfo {author}
  {\bibfnamefont {H.~P.}\ \bibnamefont {Urbach}}, \ and\ \bibinfo {author}
  {\bibfnamefont {X.}~\bibnamefont {Yuan}},\ }\href@noop {} {\bibfield
  {journal} {\bibinfo  {journal} {Nano Lett.}\ }\textbf {\bibinfo {volume}
  {14}},\ \bibinfo {pages} {6430} (\bibinfo {year} {2014})}\BibitemShut
  {NoStop}%
\bibitem [{\citenamefont {Tong}\ \emph {et~al.}(2009)\citenamefont {Tong},
  \citenamefont {Miljkovic},\ and\ \citenamefont {Kall}}]{tong2009alignment}%
  \BibitemOpen
  \bibfield  {author} {\bibinfo {author} {\bibfnamefont {L.}~\bibnamefont
  {Tong}}, \bibinfo {author} {\bibfnamefont {V.~D.}\ \bibnamefont {Miljkovic}},
  \ and\ \bibinfo {author} {\bibfnamefont {M.}~\bibnamefont {Kall}},\
  }\href@noop {} {\bibfield  {journal} {\bibinfo  {journal} {Nano Lett.}\
  }\textbf {\bibinfo {volume} {10}},\ \bibinfo {pages} {268} (\bibinfo {year}
  {2009})}\BibitemShut {NoStop}%
\bibitem [{\citenamefont {Garcia-Vidal}\ \emph {et~al.}(2010)\citenamefont
  {Garcia-Vidal}, \citenamefont {Martin-Moreno}, \citenamefont {Ebbesen},\ and\
  \citenamefont {Kuipers}}]{garcia2010light}%
  \BibitemOpen
  \bibfield  {author} {\bibinfo {author} {\bibfnamefont {F.~J.}\ \bibnamefont
  {Garcia-Vidal}}, \bibinfo {author} {\bibfnamefont {L.}~\bibnamefont
  {Martin-Moreno}}, \bibinfo {author} {\bibfnamefont {T.~W.}\ \bibnamefont
  {Ebbesen}}, \ and\ \bibinfo {author} {\bibfnamefont {L.}~\bibnamefont
  {Kuipers}},\ }\href@noop {} {\bibfield  {journal} {\bibinfo  {journal} {Rev.
  Mod. Phys.}\ }\textbf {\bibinfo {volume} {82}},\ \bibinfo {pages} {729}
  (\bibinfo {year} {2010})}\BibitemShut {NoStop}%
\bibitem [{\citenamefont {Genet}\ and\ \citenamefont
  {Ebbesen}(2007)}]{genet2007light}%
  \BibitemOpen
  \bibfield  {author} {\bibinfo {author} {\bibfnamefont {C.}~\bibnamefont
  {Genet}}\ and\ \bibinfo {author} {\bibfnamefont {T.~W.}\ \bibnamefont
  {Ebbesen}},\ }\href@noop {} {\bibfield  {journal} {\bibinfo  {journal}
  {Nature}\ ,\ \bibinfo {pages} {39}} (\bibinfo {year} {2007})}\BibitemShut
  {NoStop}%
\bibitem [{\citenamefont {Lezec}\ \emph {et~al.}(2002)\citenamefont {Lezec},
  \citenamefont {Degiron}, \citenamefont {Devaux}, \citenamefont {Linke},
  \citenamefont {Martin-Moreno}, \citenamefont {Garcia-Vidal},\ and\
  \citenamefont {Ebbesen}}]{lezec2002beaming}%
  \BibitemOpen
  \bibfield  {author} {\bibinfo {author} {\bibfnamefont {H.~J.}\ \bibnamefont
  {Lezec}}, \bibinfo {author} {\bibfnamefont {A.}~\bibnamefont {Degiron}},
  \bibinfo {author} {\bibfnamefont {E.}~\bibnamefont {Devaux}}, \bibinfo
  {author} {\bibfnamefont {R.~A.}\ \bibnamefont {Linke}}, \bibinfo {author}
  {\bibfnamefont {L.}~\bibnamefont {Martin-Moreno}}, \bibinfo {author}
  {\bibfnamefont {F.~J.}\ \bibnamefont {Garcia-Vidal}}, \ and\ \bibinfo
  {author} {\bibfnamefont {T.~W.}\ \bibnamefont {Ebbesen}},\ }\href@noop {}
  {\bibfield  {journal} {\bibinfo  {journal} {Science}\ }\textbf {\bibinfo
  {volume} {297}},\ \bibinfo {pages} {820} (\bibinfo {year}
  {2002})}\BibitemShut {NoStop}%
\bibitem [{\citenamefont {Juan}\ \emph {et~al.}(2009)\citenamefont {Juan},
  \citenamefont {Gordon}, \citenamefont {Pang}, \citenamefont {Eftekhari},\
  and\ \citenamefont {Quidant}}]{juan2009self}%
  \BibitemOpen
  \bibfield  {author} {\bibinfo {author} {\bibfnamefont {M.~L.}\ \bibnamefont
  {Juan}}, \bibinfo {author} {\bibfnamefont {R.}~\bibnamefont {Gordon}},
  \bibinfo {author} {\bibfnamefont {Y.}~\bibnamefont {Pang}}, \bibinfo {author}
  {\bibfnamefont {F.}~\bibnamefont {Eftekhari}}, \ and\ \bibinfo {author}
  {\bibfnamefont {R.}~\bibnamefont {Quidant}},\ }\href@noop {} {\bibfield
  {journal} {\bibinfo  {journal} {Nature Phys.}\ }\textbf {\bibinfo {volume}
  {5}},\ \bibinfo {pages} {915} (\bibinfo {year} {2009})}\BibitemShut {NoStop}%
\bibitem [{\citenamefont {Ebbesen}\ \emph {et~al.}(1998)\citenamefont
  {Ebbesen}, \citenamefont {Lezec}, \citenamefont {Ghaemi}, \citenamefont
  {Thio},\ and\ \citenamefont {Wolff}}]{ebbesen1998extraordinary}%
  \BibitemOpen
  \bibfield  {author} {\bibinfo {author} {\bibfnamefont {T.~W.}\ \bibnamefont
  {Ebbesen}}, \bibinfo {author} {\bibfnamefont {H.~J.}\ \bibnamefont {Lezec}},
  \bibinfo {author} {\bibfnamefont {H.~F.}\ \bibnamefont {Ghaemi}}, \bibinfo
  {author} {\bibfnamefont {T.}~\bibnamefont {Thio}}, \ and\ \bibinfo {author}
  {\bibfnamefont {P.~A.}\ \bibnamefont {Wolff}},\ }\href@noop {} {\bibfield
  {journal} {\bibinfo  {journal} {Nature}\ }\textbf {\bibinfo {volume} {391}},\
  \bibinfo {pages} {667} (\bibinfo {year} {1998})}\BibitemShut {NoStop}%
\bibitem [{\citenamefont {Degiron}\ \emph {et~al.}(2004)\citenamefont
  {Degiron}, \citenamefont {Lezec}, \citenamefont {Yamamoto},\ and\
  \citenamefont {Ebbesen}}]{degiron2004optical}%
  \BibitemOpen
  \bibfield  {author} {\bibinfo {author} {\bibfnamefont {A.}~\bibnamefont
  {Degiron}}, \bibinfo {author} {\bibfnamefont {H.~J.}\ \bibnamefont {Lezec}},
  \bibinfo {author} {\bibfnamefont {N.}~\bibnamefont {Yamamoto}}, \ and\
  \bibinfo {author} {\bibfnamefont {T.~W.}\ \bibnamefont {Ebbesen}},\
  }\href@noop {} {\bibfield  {journal} {\bibinfo  {journal} {Optics Commun.}\
  }\textbf {\bibinfo {volume} {239}},\ \bibinfo {pages} {61} (\bibinfo {year}
  {2004})}\BibitemShut {NoStop}%
\bibitem [{\citenamefont {Taflove}\ and\ \citenamefont
  {Hagness}(2005)}]{taflove2005computational}%
  \BibitemOpen
  \bibfield  {author} {\bibinfo {author} {\bibfnamefont {A.}~\bibnamefont
  {Taflove}}\ and\ \bibinfo {author} {\bibfnamefont {S.~C.}\ \bibnamefont
  {Hagness}},\ }\href@noop {} {\emph {\bibinfo {title} {Computational
  electrodynamics}}}\ (\bibinfo  {publisher} {Artech House},\ \bibinfo {year}
  {2005})\BibitemShut {NoStop}%
\bibitem [{\citenamefont {Johnson}\ and\ \citenamefont
  {Christy}(1972)}]{johnson1972optical}%
  \BibitemOpen
  \bibfield  {author} {\bibinfo {author} {\bibfnamefont {P.~B.}\ \bibnamefont
  {Johnson}}\ and\ \bibinfo {author} {\bibfnamefont {R.-W.}\ \bibnamefont
  {Christy}},\ }\href@noop {} {\bibfield  {journal} {\bibinfo  {journal} {Phys.
  Rev. B}\ }\textbf {\bibinfo {volume} {6}},\ \bibinfo {pages} {4370} (\bibinfo
  {year} {1972})}\BibitemShut {NoStop}%
\bibitem [{\citenamefont {Borghese}\ \emph {et~al.}(2007)\citenamefont
  {Borghese}, \citenamefont {Denti},\ and\ \citenamefont
  {Saija}}]{borghese2007scattering}%
  \BibitemOpen
  \bibfield  {author} {\bibinfo {author} {\bibfnamefont {F.}~\bibnamefont
  {Borghese}}, \bibinfo {author} {\bibfnamefont {P.}~\bibnamefont {Denti}}, \
  and\ \bibinfo {author} {\bibfnamefont {R.}~\bibnamefont {Saija}},\
  }\href@noop {} {\emph {\bibinfo {title} {Scattering from model nonspherical
  particles: Theory and applications to environmental physics}}}\ (\bibinfo
  {publisher} {Springer Verlag},\ \bibinfo {year} {2007})\BibitemShut {NoStop}%
\bibitem [{\citenamefont {Washizu}\ \emph {et~al.}(1992)\citenamefont
  {Washizu}, \citenamefont {Shikida}, \citenamefont {Aizawa},\ and\
  \citenamefont {Hotani}}]{washizu1992orientation}%
  \BibitemOpen
  \bibfield  {author} {\bibinfo {author} {\bibfnamefont {M.}~\bibnamefont
  {Washizu}}, \bibinfo {author} {\bibfnamefont {M.}~\bibnamefont {Shikida}},
  \bibinfo {author} {\bibfnamefont {S.-I.}\ \bibnamefont {Aizawa}}, \ and\
  \bibinfo {author} {\bibfnamefont {H.}~\bibnamefont {Hotani}},\ }\href@noop {}
  {\bibfield  {journal} {\bibinfo  {journal} {IEEE Trans. Ind. Appl.}\ }\textbf
  {\bibinfo {volume} {28}},\ \bibinfo {pages} {1194} (\bibinfo {year}
  {1992})}\BibitemShut {NoStop}%
\bibitem [{\citenamefont {Wang}\ \emph {et~al.}(2009)\citenamefont {Wang},
  \citenamefont {Schonbrun},\ and\ \citenamefont
  {Crozier}}]{wang2009propulsion}%
  \BibitemOpen
  \bibfield  {author} {\bibinfo {author} {\bibfnamefont {K.}~\bibnamefont
  {Wang}}, \bibinfo {author} {\bibfnamefont {E.}~\bibnamefont {Schonbrun}}, \
  and\ \bibinfo {author} {\bibfnamefont {K.~B.}\ \bibnamefont {Crozier}},\
  }\href@noop {} {\bibfield  {journal} {\bibinfo  {journal} {Nano letters}\
  }\textbf {\bibinfo {volume} {9}},\ \bibinfo {pages} {2623} (\bibinfo {year}
  {2009})}\BibitemShut {NoStop}%
\bibitem [{\citenamefont {Min}\ \emph {et~al.}(2013)\citenamefont {Min},
  \citenamefont {Shen}, \citenamefont {Shen}, \citenamefont {Zhang},
  \citenamefont {Fang}, \citenamefont {Yuan}, \citenamefont {Du}, \citenamefont
  {Zhu}, \citenamefont {Lei},\ and\ \citenamefont {Yuan}}]{min2013focused}%
  \BibitemOpen
  \bibfield  {author} {\bibinfo {author} {\bibfnamefont {C.}~\bibnamefont
  {Min}}, \bibinfo {author} {\bibfnamefont {Z.}~\bibnamefont {Shen}}, \bibinfo
  {author} {\bibfnamefont {J.}~\bibnamefont {Shen}}, \bibinfo {author}
  {\bibfnamefont {Y.}~\bibnamefont {Zhang}}, \bibinfo {author} {\bibfnamefont
  {H.}~\bibnamefont {Fang}}, \bibinfo {author} {\bibfnamefont {G.}~\bibnamefont
  {Yuan}}, \bibinfo {author} {\bibfnamefont {L.}~\bibnamefont {Du}}, \bibinfo
  {author} {\bibfnamefont {S.}~\bibnamefont {Zhu}}, \bibinfo {author}
  {\bibfnamefont {T.}~\bibnamefont {Lei}}, \ and\ \bibinfo {author}
  {\bibfnamefont {X.}~\bibnamefont {Yuan}},\ }\href@noop {} {\bibfield
  {journal} {\bibinfo  {journal} {Nature communications}\ }\textbf {\bibinfo
  {volume} {4}} (\bibinfo {year} {2013})}\BibitemShut {NoStop}%
\bibitem [{\citenamefont {Garcia~de Abajo}(2002)}]{garcia2002light}%
  \BibitemOpen
  \bibfield  {author} {\bibinfo {author} {\bibfnamefont {F.}~\bibnamefont
  {Garcia~de Abajo}},\ }\href@noop {} {\bibfield  {journal} {\bibinfo
  {journal} {Opt. Express}\ }\textbf {\bibinfo {volume} {10}},\ \bibinfo
  {pages} {1475} (\bibinfo {year} {2002})}\BibitemShut {NoStop}%
\bibitem [{\citenamefont {Gorodetski}\ \emph {et~al.}(2013)\citenamefont
  {Gorodetski}, \citenamefont {Drezet}, \citenamefont {Genet},\ and\
  \citenamefont {Ebbesen}}]{gorodetski2013generating}%
  \BibitemOpen
  \bibfield  {author} {\bibinfo {author} {\bibfnamefont {Y.}~\bibnamefont
  {Gorodetski}}, \bibinfo {author} {\bibfnamefont {A.}~\bibnamefont {Drezet}},
  \bibinfo {author} {\bibfnamefont {C.}~\bibnamefont {Genet}}, \ and\ \bibinfo
  {author} {\bibfnamefont {T.~W.}\ \bibnamefont {Ebbesen}},\ }\href@noop {}
  {\bibfield  {journal} {\bibinfo  {journal} {Phys. Rev. Lett.}\ }\textbf
  {\bibinfo {volume} {110}},\ \bibinfo {pages} {203906} (\bibinfo {year}
  {2013})}\BibitemShut {NoStop}%
\end{thebibliography}%

\end{document}